# Challenges in migrating legacy software systems to the cloud—an empirical study

Mahdi Fahmideh[1], Farhad Daneshgar[1], Ghassan Beydoun[2], Fethi Rabhi [1]
1 University of New South Wales, Sydney, Australia
2 University of Technology Sydney, Australia

**Abstract**

Moving existing legacy systems to cloud platforms is a difficult and high cost process that may involve technical and non-technical resources and challenges. There is evidence that the lack of understanding and preparedness of cloud computing migration underpin many migration failures in achieving organisations' goals. The main goal of this article is to identify the most important challenging activities for moving legacy systems to cloud platforms from a perspective of reengineering process. Through a combination of a bottom-up and a top-down analysis, a set of common activities is derived from the extant cloud computing literature. These are expressed as a model and are validated using a population of 104 shortlisted and randomly selected domain experts from different industry sectors. We used a Web-based survey questionnaire to collect data and analysed them using SPSS Sample T-Test. The results of this study highlight the most important and critical challenges that should be addressed by various roles within a legacy to cloud migration endeavour. The study provides an overall understanding of this process including common occurring activities, concerns and recommendations. In addition, the findings of this study constitute a practical guide to conduct this transition. This guide is platform agnostic and independent from any specific migration scenario, cloud platform, or an application domain.

## 1. Introduction

Cloud computing platforms, like those offered by Amazon Web Services, AT&T, GoGrid, Rackspace Cloud Hosting, HP, Yahoo, Intel, IBM, and Google cloud have received significant attention in addressing the requirements of IT-based organisations such as computational power, reducing the cost of infrastructure and maintenance, and efficient resource usages [1-3]. These benefits are realised by a wide range of services which are universally accessible, dynamically acquirable and releasable, and through usage based costing models. Cloud computing is a fundamental shift in delivering IT services with broad impact on organisations. Cloud enablement of existing legacy systems, as a backbone of organisations, has significant implications for organizational performance. A recent report by Gartner concludes that, by 2020, an organisation with *No-Cloud* policy will be as rare as one with *No-Internet* policy is today [4]. This momentum is concomitant to an ever increasing number of dedicated journal special issues, conferences, conference tracks, and workshops on cloud migration. These trends provide the backdrop of this paper.

The industry and academia discussions on the cloud computing technology vary from pure technical solutions for utilizing cloud services to theoretical analysis of its potential socio-technical impact. According to a literature review conducted in [5] and elaborated later in [6], software engineering and practitioners' reports typically suggest partial or technical solutions applicable to only a specific stage of the cloud migration process. The current body of cloud computing knowledge still lacks detailed and overarching view to reengineer legacy systems to make them cloud-enabled. This requires tying together these bins and ad-hoc solutions fragmented in the literature to understand how cloud migration process can be systematically conducted. That is, organisations often lack a complete understanding of circumstances to successfully undertake the complex changes required in their legacy systems to attain advantages of cloud services [7]. We are not alone in observing this.

Mahmood [8] points out "although the technologies underlying the cloud paradigm have been in existence for some time, the frameworks and methodologies for construction, deployment and delivery of cloud services and environments are in the process of being further developed and refined" (p. viii). Calls have also been made to focus further studies on exploration of what a typical cloud transition entails along with relevant concerns for taking into account [5, 9, 10]. Notwithstanding, this recent surge of awareness, not much progress has been made as a recent survey by Fahmideh et. al [6] highlights.

Further research can certainly facilitate identification of critical difficulties and their organisation, anticipation of events, cost involved, and expected outcomes from the legacy system to cloud enablement. This can help IS executives to carry out a safe and risk-aware migration, instead of struggling to identify what and how to move their existing legacy systems to the cloud or moving from an existing cloud platform to another one in the post-migration phase in an ad-hoc manner leading to a poor, unsecured migration, and a cost blowout. Our study attempts to address the call made by [6] on investigating how legacy systems are reengineered to utilize cloud services, critical activities incorporating, and ways they are conducted. We explore this process in-depth and uncover critical activities, work-products, concerns, and recommendations along with getting insights from domain experts. This study concentrates on the following research question: What are the key critical difficulties that are incorporated into the cloud migration process and should be systematically addressed? In this research, legacy systems are viewed as the unit of analysis, a socio-technical IT artefact embedding in business processes of organisations.

The remaining parts of the paper continue with Section 2 where a discussion is provided on related work and how our study is situated in the context of the current literature. Section 3 explains the research methodology used to answer the research question of the study. Section 4 describes a derived model for moving legacy systems to the cloud along with domain experts' suggestions. Finally, this paper concludes with a discussion on research implications of the current study and its further directions in sections 5 and 6, respectively.

## 2. Prior Research

### 2.1 Legacy systems

The term *legacy system* firstly appeared in IT literature in 1990 meaning "relating to, or being a previous or outdated computer system" [11]. Many other definitions can be found. For instance, an earlier definition, from [12], says "large software systems that we don't know how to cope with but they are vital to our organisation" (p.19). Holland [13] mentions that "legacy applications encapsulate the existing business processes, organization structure, culture, and information technology" (p.31). Studies [14, 15] characterise legacy systems as those that are mission critical, expensive to maintain, brittle and inflexible to changes, run on obsolete hardware, incomplete or outdated documentation, and difficult to extend and integrate with other systems. The term legacy system is often co-located with old mainframe-based systems that are written in programming languages such as FORTRAN and COBOL and processing users' transactions. However, modern systems such as web-based systems which have been developed using the latest technologies available in the marketplace, but currently do not satisfy new business requirements are also considered as legacy system [16, 17]. A common impression provided in all above definitions is the worthiness of the legacy systems and this has been the reason to keep them working in organisations. Legacy systems support business processes, maintain organisational knowledge, and provide significant competitive advantage with a positive return and contribution to the organisation revenue and growth [12, 18, 19].

After few decades of research on legacy system modernisation, there are still many operational legacy systems continuously evolving to support IT functions of organizations. They are characterised by the need for high computing capability, reducing the cost of maintenance and upgrading, efficient resource utilisation, and less environmental impact and electrical energy consumption [1-3]. In addressing these modernisation requirements, cloud computing has been recognised as a promising initiative. The primary aim of legacy system migration to the cloud is essentially to reduce maintenance cost, increase reusability, and extend their functionalities [6].

## 2.1 Cloud Migration Research

In the current research, we distinguish between IT adoption and IT migration because migration happens from an incumbent IT to a substitute IT, whereas adoption does not necessarily assume an existing IT [20]. Exploring cloud migration as an instance of switching from client-hosted systems to cloud-hosted systems is an on-going research track and has been studied in the literature from a variety of perspectives including models for organisations to decide on the suitability of cloud migration [21], tools for making decision about selecting suitable cloud services [22], inhibitors and enablers associated with this cloud technology [23], its benefits in terms of enhanced competitive advantages [24], standardization [25], and interoperability issues [26], to mention a few. For example, drawn from a sample of students using Google Apps, Bhattacherjee et al. in [20] examined why end users migrate from client-hosted to cloud computing platforms. The study presents and empirically validates a model showing pull factors (e.g. dissatisfaction with client IT and relative usefulness) and push factors (e.g. learning cost, setup cost, and security concerns). However, the authors state that despite the presence of these factors, users may not universally migrate to the cloud due to obstacles such as switching costs and personal judgment. In addition, in [27], the agency theory has been used to understand and analyse conflicts of interests occurring between cloud service providers and consumers. The study categorizes and assesses countermeasures to these conflicts and to identify possible technical and non-technical means to extend the practicality of cloud migration. Using the transaction cost theory, another study, [28] observed that organisations have concerns about the cloud services because liabilities may not be clearly specified as traditional outsourcing do and also the nature of cloud services involve different roles and foreign legislation. Such uncertainty appears in the form of opportunism of cloud service provider, confidentiality, integrity, and availability of information. The above research views the cloud migration as a 'black box' by not narrowing its focus down on key internal operational difficulties involved in the migration process.

In the software engineering literature majority of studies that investigate cloud migration are mainly technically oriented with focus on: reusing legacy system codes and deploying them in the cloud [29], addressing interoperability issues [30], and optimum distributions of legacy system components on cloud servers (e.g. CloudGenius) using applying different techniques such as simulation [31, 32]. None of the existing studies provide a holistic picture of migration process nor take into consideration key challenges inherent to the cloud migration, such as multitenancy or scalability, nor provides an empirical validation. We have yet to see suggestive evidence of key challenges when legacy systems are moved from existing on-premise local-hosted systems to cloud-hosted systems for their data, functionalities, and storage needs. This constitutes the main motivation of the current study.

A few definitions of legacy system to cloud migration process exist in the literature. Chauhan at el. [33] define this process as a kind of a reengineering process to enable interaction with and utilization of cloud services. Another definition views migration as a re-architecting process applying a set of modifications on different components (business and data layers) of legacy systems in order to make them cloud complaint [34]. A broad definition encompassing socio-technical aspects of the cloud migration defines this process as "a set of migration activities carried out to support an end-to-end cloud migration. Cloud migration processes define a comprehensive perspective, capturing business and technical concerns of Stakeholders with different backgrounds are involved" [35]. But, it is still remain unanswered question as what key critical and unique activities are involved in legacy system to cloud migration in comparison to traditional legacy system reengineering.

Migrating large-scale legacy systems is a challenging exercise since they often predate cloud computing and may have been developed without taking into account the unique characteristics of cloud environments [6] and [34]. For example, the cloud environment is characterised with a an infinite pool of resources such as CPU, memory, storage, and network bandwidth which can be acquired and released by service consumers, based on demand to optimise resource usage in the case of fluctuation in workload. This characteristic is often referred to elasticity that distinguishes this computing paradigm from other ones such as cluster and grid computing [36]. However, some legacy systems might not have been implemented with a support of dynamic scaling up/down of cloud resources; rather they assume that the elasticity is supported by providing more powerful physical servers. As such, the legacy system architecture should be refactored and modified to support the feature of elasticity. In addition, the literature indicates that the migration process has sometimes failed to achieve organisations' goals due to: (i) inadequate understanding of cloud computing requirements, (ii) early engagement with technical implementation phase of cloud enablement, (iii) lack of planning, (iv) being simply seduced by magazine/hype, and (v) facing unexpected issues that were out of the control of service consumers and providers [37-40]. A major concern of an organisation is how to avoid failure in cloud environment. For example, Google's email service failure caused many user's emails, themes, folders, and personalized settings be erased [41].

Taking into account the above issues, evidence suggests that not all organisations are rushing to utilize cloud services [42, 43]. Due to these difficulties, some high-privacy and safety critical systems such as military, aviation, and aerospace systems might not be able to directly take advantage of cloud services [34, 44]. Others such as business information systems can be reengineered to utilise cloud services and thus are adaptable to the cloud platforms. Understanding the nature of moving such systems to cloud platforms is the focus of the current research.

Typically, software systems constitute several layers mainly a user interface, business logic, and data layers along with databases. Each layer may have several components deployable on different servers and work collaboratively. According to [6], there is a few possibilities that these layers of a system can become cloud enabled based on the common type of cloud service delivery models IaaS, PaaS, and SaaS. They are called migration variants and are defined as follow:

- *Type I*: Deploying business logic layer of legacy systems, offering independent and reusable functionalities, on the cloud server/infrastructure by applying the service delivery model IaaS. However, the data associated with the legacy systems are kept in local organisational network. Deploying an image-processing component of a legacy system on Amazon EC2, is an example of this migration type.
- *Type II*: Re-engineering a legacy system to SaaS or replacing some of its components with an available and fully tested SaaS. The Salesforce CRM system is a typical example of SaaS, which can be integrated with other legacy systems via its interfaces or wrapping mechanisms.
- *Type III*: Deploying the legacy database on a cloud data store provider through IaaS delivery service model. In this migration type, the components related to business logic layer are kept in local organisational network and the database is deployed on a cloud data store such as Amazon Simple Storage Service (S3), Amazon Elastic Block Store (EBS), Dropbox, Zip Cloud, and Just Cloud.
- *Type IV*: Converting the data, schema, and modifying the data of a legacy system to a cloud database solution provider such as Amazon SimpleDB, Google App Engine data store, or Google Cloud SQL.
- *Type V*. Deploying the whole legacy system stack on a cloud infrastructure via service delivery model IaaS. The legacy system is encapsulated into a single Virtual Machine (VM) and then is run on the cloud infrastructure. Hosting of a Web application and its Web server as a VM on EC2 is an example of such a migration variant.

The reason for selecting a particular or a combination of migration variants can be due to security, network performance, or migration cost. These variants will be used in Sections 3.1 to show how different cloud migration scenarios may entail different requirements.

The current study adds to the body of knowledge on the cloud migration by providing a consolidated understanding of the legacy to cloud migration process which supersedes previous work in the following ways. Unlike previous research which tends to explore cloud enablement of IT-based organizations as a unit of analysis, we have established our research as an exploratory study, aiming to narrow down our view and discover insights of cloud-enablement of legacy systems. In addition, we not only identify key important technical and non-technical relevant critical process elements and corresponding recommendations associated with such transition, but also we empirically confirm them by getting insights from experts to further increase the reliability of the findings. Furthermore, our study is a response to the call made by [6] to empirically investigate how reengineering process of legacy systems to the cloud is conducted; and hence the process-based view of this transition is provided for both IS researchers and practitioners as a guidance.

# 3. Research methodology

Given the objective of our study and in the light of our literature review, we used a mixed-methods approach in our attempt to provide an exhaustive understanding of the legacy to cloud migration and to identify key process elements involved in this transition [45]. There are some studies that have concurrently taken advantage of both qualitative and quantitative methods such as the work by [46] on customer-oriented citizenship behaviours of IS professionals, on the exploration of technical debt in IS [47], on the role of behavioural control on trust decline in virtual teams [48], and analysing effective mechanisms in cross-project learning [49]. The qualitative and quantitative methods are complementary to each other and can lead to a richer understanding of the phenomena and increase reliability and accuracy of results [50]. We followed Venkatesh's guidelines [51] in designing mixed research method. Table 1 shows how these guidelines are mapped to the context of this research. We conducted a two-phase research process, each step employing both qualitative and quantitative methods in different forms. In the first phase, we derived a model showing the key elements that a generic cloud process entails. This probing model can be used by organizations to examine the extent to which these process elements are in place. This is further outlined in Section 4. In the second phase, we report on the structured survey results of cloud computing experts to illustrate how the various items identified in the process model come to real-world cloud migration scenarios and thus provide a rich description of these elements.

Table 1. Guidelines by [51] for conducting the mixed-methods

| Guideline | Evaluation criteria | Specialization in this study |
|---|---|---|
| Decision on the appropriateness of a mixed-methods approach | Consider the core research objective and check if the mixed-methods is suitable for an inquiry | We used the mixed-methods to assess the credibility of our understanding about what legacy system migration to cloud environments entails from the perspective of process. Based on an intensive qualitative analysing of existing anecdotes in the literature, a conceptual model of the cloud migration process was derived. We confirmed the relevancy and soundness of this process model using a quantitative survey and qualitative feedback from domain experts. |
| Define a strategy for mixed-methods research design | Evaluate the mixed methods from the perspectives of research objectives and sough theoretical contributions | Qualitative and quantitative data collections were undertaken in this research. Firstly, qualitative data collection, i.e. anecdotal descriptions of cloud migration process in the existing literature, enabled us to understand how the process of transition to cloud environments is defined by the literature. Next, the quantitative and quantitative analysis of the survey data allowed us to confirm our findings, i.e. the process model derived from the literature. |
| Develop a strategy for analysing mixed-methods data | Apply norms in qualitative and quantitative analysis techniques that are commonly used for developing and evaluating models | For qualitative analysis in phase 1, we followed common qualitative data analysis techniques for the metamodeling including extracting and short-listing a set of candidate elements from the existing sources, reconciling their definitions, organizing elements into migration phases, and classifying elements based on the migration variants (Section 2.1). For phase 2, we performed One Sample T-Test to confirm soundness of the critical elements in the process model. Furthermore, complementary qualitative comments provided by domain experts were used to show why elements in the model are perceived as important for consideration in a real-world scenario of legacy system to cloud migration. |
| Make meta-inferences from the results arising from the mixed-methods analysis | Meta-inferences should be based on the viewpoint of research objective | In section 4 (Findings), we provided a detailed description of the cloud migration process by triangulating results from the phases 1 and 2. |

## 3.1 Phase 1— Derivation of key critical challenges of cloud migration process

The first phase was the derivation of a set of key critical activities during cloud migration process synthesized from the existing literature. In this research, we viewed *anecdotal descriptions* of challenges involving legacy system to cloud migration in the literature in order to derive a generic view of cloud migration process model. Anecdotes have been recognized as important sources of information for understating the nature of technology adoption, particularly in applied area. For example studies by [52] on developing a process model of IS flexibility, [53] on service-oriented system development, [54] on model-driven disaster management, and [55, 56] designing a metamodel of agent-based system development all use anecdotal description. This is also aligned with the recommendation by Schwarz, who states that a conceptual model *synthesizes previous research in an actionable way for practitioners* [57]. In this research, this phase was performed by extensive qualitative analysis of existing papers in the cloud migration literature. For the data analysis we adopted the following steps of thematic analysis as per the guidelines in [55, 56, 58]:

**(i) Identifying studies.** We first identified all relevant reports (i.e. case studies, experience reports) on the legacy system migration to cloud platforms. Every year a considerable number of research papers have been published, each reporting different set of successful experience and challenges. As a result, recommendations in conducting a systematic literature review [59] seemed an appropriate review method for the current study to ensure all major relevant studies are included in the derivation of the target process model. These recommendations included (i) defining search strings, (ii) selecting study sources, (iii) defining study selection criteria, (iv) conducting the review, and (v) extracting demographic data.

**(ii) Extracting and shortlisting key activities, work-products, and general principles.** In this research an item refers to one of the following (i) *activity*: a discrete and small unit of migration work that IT developers may perform to achieve one or more specified goal, (ii) *work-product*: a tangible artefact that is produced during the migration process and used by other activities, and (iii) *design principles* that are taking into account during a cloud architecture design. We first carefully reviewed all identified papers and then specific segments of text related to an item were stored in a table, along with its names and references. The following criteria were used to identify and extract items from the papers: (i) an element should be sufficiently generic to a variety of cloud migration scenarios regardless of a particular cloud platform and underlying technology (ii) an item should be cloud-specific and related to one or more concerns relevant to the cloud migration process mainly for analysing organisational context, understanding cloud migration objectives and requirements, proper cloud migration planning, understanding legacy systems, target cloud platform/service selection, re-architecting legacy systems, environment configuration, and testing as described in [6]. Note that, items related to the post migration phase such as monitoring and optimising the legacy systems were out of the scope of the current study.

**(iii) Creating a model of higher-order themes**. By adopting a bottom-up approach, this step has focused on analysing and grouping the extracted items, analysing their similarities, classifying, and reconciling differences in order to group all related text segments, and generating a set of high-level elements that together form a view of the migration process.

**(iv) Defining relationships between migration variants and the process elements**. It is important to realise that the migration variants (i.e. I, II, III, IV, and V), presented in Section 2.1, may raise some concerns to be addressed during the migration process. In this regard, some elements in the model are deemed constant in any migration scenarios. For example, as it has also been confirmed by [60], it is necessary to analyse the impact of organisational changes, risks, and effects associated with the cloud migration before the migration process proceeds. Thus, the process element *Analyse Context* is mandatory in all types of migration variants. This is also true for the activity *Design Cloud Solution* which is to define a new architecture model/topology of migrated legacy system components on the cloud and their communication with none-migrated components in local organisational network. Derived from the studies identified in phase 1, we found the following process elements as mandatory one in any legacy system to cloud migration process: Analyse Context, Analyse Migration Requirements, Define Plan, Choose Cloud Platform/Provider, Design Cloud Solution, Cloud Solution

Architecture Model, Deploy System Component, Handle Transient Faults, Synchronise/ Replicate System Components, and Test.

On the other hand, some elements are situational, largely depending on characteristics of legacy systems, the choice of a migration variant, cloud architecture design, and cross-cutting concerns such as security, elasticity, performance variability, network latency, and availability. For example, due to some incompatibility issues such as APIs, interfaces or data types between legacy systems and cloud services, it might be required to perform some modifications in the source codes of legacy systems. This situation is possible to occur when all migration variants (Section 2.1) are applied, except for migration type V [34, 61].

**(v) Assess the trustworthiness of the synthesis**. To ensure the reliability of the findings, we applied peer crosschecking as a technique among the authors of current article. This resulted in an intense and fruitful discussion among the authors involved in this study as to whether the elements of the model correctly capture the essence of the reviewed literature. A full description of conducting the abovementioned steps is publically available upon request.

**Resultant model**. The model in Figure 1 shows a typical legacy to cloud transition process without situation-specific operationalisation and technical details. This yields in an abstract view of what a transition process may entail. Table 2 shows all situations in which an element should be incorporated into the migration process with respect to a given migration variant. In this table, symbols √, (√), and × specify a *Mandatory*, *Situational*, and *Unnecessary* adoption of an element, respectively.

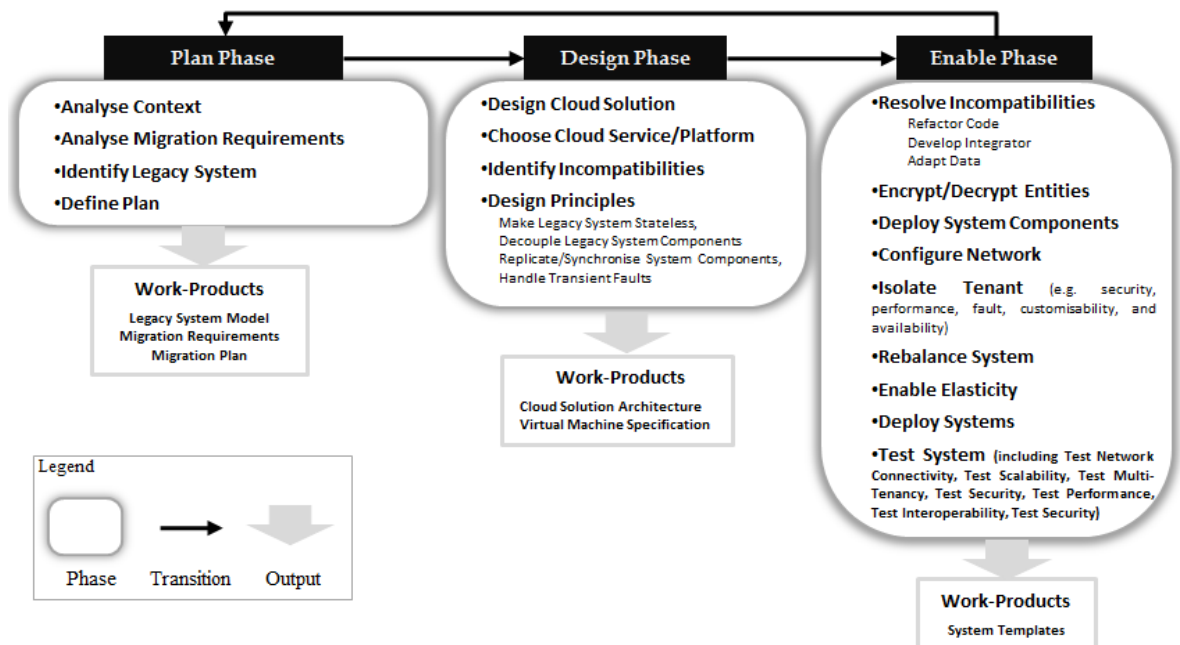

Figure 1. Critical process elements in moving legacy systems to cloud platforms

Table 2. Key elements incorporating into the legacy system migration to cloud platforms systems

| Process element | Definition | Goal | Migration Type | | | | |
|---|---|---|---|---|---|---|---|
| | | | I | II | III | IV | V |
| Analyse Context | Analyse migration suitability with respect to factors such as cost of system modification, installation, training, administration, license management, required expertise, pricing models of the service providers, infrastructure procurement imposed by the migration, impact of the cloud on stakeholders, organisational constraints, responsibilities, and working practices. | Cost estimation, risk mitigation | √ | √ | √ | √ | √ |
| Analyse Migration Requirements | Identify a set of requirements to be satisfied by the cloud such as computational requirements, servers, data storage and security, network and response time, and elasticity. | Project management, quality assurance, traceability to migration requirements | √ | √ | √ | √ | √ |
| Choose Cloud Platform/Provider | Define a set of suitability criteria that characterise desirable features of cloud providers including pricing model, constraints, offered QoS, electricity costs, power and cooling costs, organisation migration characteristics (migration goals, available budget), and system requirements. | Finding a best possible providers that address the migration requirements | √ | √ | √ | √ | √ |
| Decouple Legacy System Components | Decouple legacy system components from each other. Use mediator and synchronisation mechanisms to manage interaction between the loosely coupled components. | Facilitating scalability, increasing maintainability, increasing transparency | (√) | (√) | (√) | (√) | (√) |
| Define Plan | Define a correct and safe sequence of activities which guide the migration process by analysis feedback from stakeholders. A plan may include (i) notice of temporal unavailability of legacy systems, (ii) roll-back to in-house versions, (iii) migration variant and the degree of migration (.i.e. complete or partial), and (iv) retirement procedures. | Project management, risk mitigation | √ | √ | √ | √ | √ |
| Deploy System Component | Install system components and any required third party tools in the cloud. | Make the system operational | √ | √ | √ | √ | √ |
| Design Cloud Solution | Identify those legacy systems which are appropriate for the migration regarding migration requirements and then define their distribution on cloud servers. | Finding an optimum set of legacy parts for the cloud migration and their distribution in the cloud. | √ | √ | √ | √ | √ |
| Develop Integrators | Develop mediators/wrappers to hide incompatibilities occurring at runtime between legacy system components and cloud services that are plugged to these system components. | Increasing interoperability, Reducing vendor lock-in, Increasing transparency access to the system components in the cloud | (√) | (√) | (√) | (√) | (√) |
| Enable Elasticity | Define scaling rules and provide support for dynamic acquisition and release of cloud resources. | Increasing resource efficiency | (√) | (√) | × | × | (√) |
| Encrypt/Decrypt Database | Encrypt critical databases prior to hosting them on the cloud. | Increasing security | × | (√) | (√) | (√) | (√) |
| Encrypt/Decrypt Messages | Secure messages transmission between the local components and those hosted on the cloud or distributed across multiple clouds using an encryption mechanism. | Increasing security | (√) | (√) | (√) | (√) | (√) |
| Handle Transient Faults | Detect and handle transient faults may occur in the cloud. | Handling accidental faults | √ | √ | √ | √ | √ |

| Task | Description | Benefits | | | | | |
|---|---|---|---|---|---|---|---|
| Identify Incompatibilities | Identify technical mismatches between legacy system components and selected cloud service. | An estimation of cost and effort to resolve incompatibilities | √ | √ | × | √ | √ |
| Identify Legacy System | Produce a complete representation of legacy system architecture including its data, components, dependencies among components and infrastructure, system data usage and resource utilisation model (e.g. CPU, Network, storage). | Understanding of the legacy system dependencies | √ | √ | √ | √ | √ |
| Isolate Tenant | Protect tenants' data security, performance, customizability, availability, and faults from other tenants which are running on the same cloud server. | Reducing fault resiliency, Increasing availability, keeping tenant performance, increasing security | × | √ | × | × | × |
| Make Legacy System Stateless | Enable legacy system to handle safety and traceability of tenant's session when various system instances hosted on the cloud. | Facilitating scalability, increasing maintainability, increasing transparency | (√) | (√) | (√) | (√) | (√) |
| Make Mock Migration | Build a prototype of cloud solution to get an understanding of how the functional and none-functional aspects of the system will work on the cloud. | Testing quality of system architecture, an early approximation of a final cloud-enabled legacy system | (√) | (√) | (√) | (√) | (√) |
| Obfuscate Codes | Protect unauthorised access to code blocks of system components by other tenants that are running on the same cloud provider. | Increasing security | (√) | (√) | × | × | (√) |
| Re-configure Network | Re-configure the running environment of the system including reachability policies to resources and network, connection to storages, setting ports and firewalls, and load balancer. | Increasing security | √ | √ | √ | √ | √ |
| Replicate/ Synchronise System Components | Deploy system components (e.g. database, business logic) on multiple cloud servers and provide a support for synchronising them, .i.e. replica, hosted on on-premise network and cloud servers. | Guarantee consistency of system components, increasing availability | (√) | (√) | (√) | (√) | (√) |
| Test Interoperability | Test compatibility of components with different cloud environments specifically, when components can be switched between different infrastructures. | Detecting interoperability issues | (√) | (√) | × | (√) | (√) |
| Test Multi-tenancy | Test if tenants can easily configure the system components, i.e. user interfaces, business logic and workflows, and functional services. | Detecting multi-tenancy issues | × | √ | × | × | × |
| Test Network Connectivity | Test the network connectivity between local components and those hosted on the cloud. | Assuring connectivity between local-hosted components and cloud-hosted components | (√) | (√) | (√) | (√) | (√) |
| Test Performance | Test the performance (e.g. process speed, response time, throughput, latency, and etc.) of the system when subjected to increased load from multiple tenants. | Detecting performance bottlenecks | (√) | (√) | (√) | (√) | (√) |
| Test Scalability | Test to assure the system acquire and release the computing resources in an efficient manner. | Detecting scalability issues | (√) | (√) | × | × | (√) |
| Test Security | Test system components and reachability policies to access these components against the security requirements. | Detecting security issues | (√) | (√) | (√) | (√) | (√) |

## 3.2 Phase 2 — Survey

The purpose of the second phase was to examine if the process elements derived in the previous phase are perceived as sound and relevant for consideration in reengineering of legacy systems to cloud platforms. We followed the guidelines for survey design, sampling, and data collection outlined in [62].

**Survey design and data collection**. A Web-based survey was designed using Qualtrics [63], which is a Web-based survey designer software and is able to store survey results. Questions related to each process element were derived from the studies identified in the phase 1 and their definitions in Table 2. Responses to the survey questions fall into a spectrum of seven scales (1-7) where 1 represents 'completely irrelevant', 2 indicates 'unimportant', 3 for 'somewhat unimportant', 4 for 'neither important nor unimportant', 5 for 'somewhat important', 6 for 'important', and 7 for 'extremely important'.

The survey was public and distributed via sending email to cloud computing experts who had a profile in the social media such as Twitter, Linkedin, Facebook, and academic research groups. To increase the credibility of the collected responses, purposeful sampling technique [64] was used. The target eligible population for participation in the survey was decided to be those who had real-life experience in cloud migration, as well as the academia who had already published papers in academic journals and conference proceedings related to the cloud computing. Prior to the recruitment of each participant and through an inquiry email, we questioned each participant if he/she had real-life experience and/or high-level academic knowledge on cloud related process activities that may have not been reflected in their social media profile. If their response and profile combined were satisfactory, then the survey and an invitation letter would be sent to them through a second email. Furthermore, to extend the target population of respondents, the snowball sampling method [65] was employed meaning that the first identified respondent helped us to identify other respondents who could satisfy the criteria for the participation in the survey. Participation in the survey was voluntary. Close to the end of the data collection period, a friendly reminder e-mail was sent to respondents who had not answered the survey. In total 515 experts were initially invited for the participation and 144 qualified respondents answered the survey. Of 144 responses, 104 were usable after excluding incomplete answers, giving a final response rate of 20%. The respondents were from 32 countries. The average of cloud migration experience of respondents was 3.8 years.

**Quantitative data analysis**. In this research, the main question facing researchers is "what are critical elements, .i.e. activities, work-products, and design principles that should be properly addressed in the migration process of legacy to cloud platforms?". Data obtained from the questionnaire were then tested using SPSS statistical package. One Sample T-Test [66] was used to examine if the importance ratings of the process elements were distributed away from a known median of 5, a test-value, representing 'somewhat important'.

Several assumptions were checked before using the above statistical test. Firstly, the normality of the sample data was checked using Kolmogorov–Smirnov test [67]. This detected that the collected data deviates from the normality assumption. Nevertheless, according to the central limit theorem [68], a sufficiently large sampling distribution, which was N=104 in our case, will approximately have a normal distribution. This argument has also been confirmed by a variety of results from existing studies that have shown parametric tests are not sensitive to the non-normality assumption [69, 70]. Secondly, One Sample T-Test assumes that variables are measured at a ratio level which was true in our case as constructs were rated using Likert scale from 1 to 7. And finally, One Sample T-Test assumes that the responses are collected independently in the sense that the respondents are randomly selected and they individually participate in the survey without any cross-effects. This assumption was also true in our data collection as no respondent was aware of the identity of the final set of respondents. Once the assumptions of One Sample T-Test were satisfied, the test was applied for statistical analysis using SPSS software [71].

**Findings**. Table 3 shows the descriptive statistic and the results of One Sample T-Test's with $p < .05$ for each element. Since the hypothesis was formulated for a right-tailed test, the $p$ values in Table 3 were divided by 2. Column six of this table shows the majority of the elements defined in the proposed process model (27 out of 29) were regarded significant by respondents. This finding confirms the relevance and importance of the defined constructs in the suggested conceptual model. On the other hand, the *Tenant Isolation* (t-test statistic of= -1.499, *p-value of*=0.06) and the *Virtual Machine Specification* (t-test statistic of=0.513, *p-value of* =0.30) were not perceived important.

Table 3. Descriptive statistics for the constructs and the results of one sample T-Test for each element

|  | Process element | Mean | Std. Deviation | T-Statistic value | *p-value* |
|---|---|---|---|---|---|
| 1) | Adapt Data | 4.6250 | 1.16742 | -3.276 | 0.000 |
| 2) | Analyse Context | 5.6827 | .97805 | 7.118 | 0.000 |
| 3) | Analyse Migration Requirements | 6.1346 | .75115 | 15.404 | 0.000 |
| 4) | Choose Cloud Platform/Provider | 5.9519 | .92830 | 10.458 | 0.000 |
| 5) | Configure Network | 5.7476 | 1.03581 | 7.325 | 0.000 |
| 6) | Decouple Legacy System Components | 5.9417 | .94791 | 10.083 | 0.000 |
| 7) | Define Plan | 6.3365 | .67710 | 20.130 | 0.000 |
| 8) | Design Cloud Solution | 6.2692 | .85025 | 15.223 | 0.000 |
| 9) | Develop Integrators | 4.4706 | .95135 | -5.620 | 0.000 |
| 10) | Deploy System | 5.2404 | 1.31086 | 1.870 | 0.032 |
| 11) | Enable Elasticity | 5.8641 | .95022 | 9.229 | 0.000 |
| 12) | Encrypt Database | 5.2692 | 1.28645 | 2.134 | 0.017 |
| 13) | Encrypt/Decrypt Messages | 5.6923 | 1.07104 | 6.592 | 0.000 |
| 14) | Handle Transient Faults | 5.6635 | 1.05766 | 6.397 | 0.000 |
| 15) | Identify Incompatibilities | 5.9712 | .93950 | 10.542 | 0.000 |
| 16) | Make Legacy System Stateless | 5.6346 | 1.19105 | 5.434 | 0.000 |
| 17) | Obfuscate Code | 4.6311 | 1.54025 | -2.431 | 0.008 |
| 18) | Rebalance System Components | 5.3786 | 1.18086 | 3.254 | 0.001 |
| 19) | Refactor Codes | 4.6505 | .85988 | -4.125 | 0.000 |
| 20) | Replicate/Synchronise System Components | 5.8462 | .93237 | 9.255 | 0.000 |
| 21) | Isolate Tenants | 4.8269 | 1.17781 | -1.499 | 0.068 |
| 22) | Identify Legacy System | 6.1346 | .98590 | 11.736 | 0.000 |
| 23) | Test System | 6.0865 | .71204 | 15.562 | 0.000 |
| 24) | System Templates | 5.3725 | 1.16824 | 3.221 | 0.001 |
| 25) | Cloud Solution Architecture | 5.9615 | 1.00410 | 9.766 | 0.000 |
| 26) | Legacy System Model | 5.7115 | 1.11192 | 6.526 | 0.000 |
| 27) | Migration Plan | 6.1058 | .82342 | 13.695 | 0.000 |
| 28) | Migration Requirements | 5.9135 | .96653 | 9.638 | 0.000 |
| 29) | Virtual Machine Specification | 5.0686 | 1.35164 | .513 | 0.3045 |

\* p-values were divided by two as t-test was one-tailed

**Qualitative data analysis**. In a parallel effort, and to increase the meaningfulness of the quantitative responses, the participants were also requested to provide their justifications for their ratings of each element that they perceived as important. From 104 received responses, 101 participants provided qualitative responses, resulting in 30 pages of text. Responses with complete sentences and sufficiently generic for a variety of cloud migration scenarios were used for further analysis and those comments entangled with technical implementations of elements were considered out of the scope of this research and were excluded from further analysis. As a result, the comments of 31 participants were found coherent and meaningful to be used as a supplementary qualitative support for the importance of critical elements in the process model. Table 4 shows those respondents who provided qualitative explanations for their ratings. The identification code E1 to E26 was assigned for each respondent to ensure anonymity.

Table 4. Details of the respondents

| Expert | Years of experience | Role(s) | Industry sector(s) |
|---|---|---|---|
| E1 | 4 | IT Developer | Health Care, Utilities |
| E2 | 6 | IT Developer | Not Stated |
| E3 | 4 | IT Developer | IT |
| E4 | 4 | IT Consultant | Finance, IT, Telecommunication Services |

| E5  | 2          | IT Developer            | IT, Utilities                                              |
|-----|------------|-------------------------|------------------------------------------------------------|
| E6  | 5          | IT Consultant           | IT, Telecommunication, Consumer Discretionary              |
| E7  | 5          | Instructor              | Financial, IT                                              |
| E8  | 8          | IT Developer            | Energy, Financial, Telecommunication Service               |
| E9  | 3          | IT Developer            | IT, Medical                                                |
| E10 | 4          | IT Developer            | Energy, IT                                                 |
| E11 | 3          | Instructor              | Not Stated                                                 |
| E12 | 4          | Instructor              | Consumer discretionary                                     |
| E13 | 5          | IT Developer            | IT, Online Advertisement                                   |
| E14 | 7          | IT Developer            | Health Care, IT                                            |
| E15 | Not Stated | IT Developer            | Not Stated                                                 |
| E16 | 5          | IT Developer            | Not Stated                                                 |
| E17 | 3          | Academic researcher     | Health Care                                                |
| E18 | 3          | IT Developer            | IT                                                         |
| E19 | 5          | IT Developer            | Financial                                                  |
| E20 | 4          | IT Developer            | IT, Telecommunication Service, Utilities, Higher Education |
| E21 | 2          | IT Developer            | Energy, Utilities                                          |
| E22 | 2          | IT Developer            | IT                                                         |
| E23 | 5          | IT Developer            | Scientific Work-Flow, IT                                   |
| E24 | 5          | IT Developer            | IT, Utilities, Consumer discretionary                      |
| E25 | 5          | IT Developer            | Financial, IT, Telecommunication Service                   |
| E26 | 5          | IT Developer, Consultant| Financial, IT                                              |

## 4. Findings

The following subsections present a detailed explanation of all key elements in the process model organised in three phases and synthesised with relevant quotes from the participants as confirmatory evidence to justify why these elements should be incorporated and adequately addressed in a migration exercise. The process elements are presented in italic.

### 4.1 Plan phase

Transition to the cloud is not only a technological improvement of existing legacy systems, but also it is a change in the way these systems hereinafter operate, provide services, and are to be maintained. Therefore, an understanding of the impact and scope of changes associated with cloud-enablement of legacy systems is an important consideration. Project stakeholders need to perform *Analyse Context* to assess the suitability of the cloud migration for empowering existing legacy systems. This essence was mentioned frequently by the participants in the survey. For instance, E2 stated that:

> "To me it does not make sense when people talk about changing something internally in an application or system, which doesn't provide business value e.g. new features or better performance. Changing something because it's the latest thing is simply a waste of money. Thus, when moving to cloud you have well defined goals. Some questions you may ask yourself are, should I add new features or change the application behaviour now that I'm going to redesign/redevelop it?"

A similar statement of performing *Analyse Context* was also mentioned by E4:

> "Cloud migrations will almost always result in significant change to (often longstanding) business processes. Many employees will begin to worry about job security - similar to outsourcing projects. Be prepared to deal with resistance in equal quantities from business users, as well as engineers and developers".

In addition, E3 referred to the *Analyse Context* as the "Pre-Migration Assessment" meaning that "whether it is actually worthwhile to conduct the migration". Also, E5 mentioned "particularly at first, there should be reason whether there is a need to move application to the cloud, as it's not a silver bullet and not all applications gain some benefits from running in the cloud".

Furthermore, E18 stated that *Analyse Context* is for "justification of choice of technologies and providers". E1 justified the importance of *Analyse Context* from a financial perspective:

"Financial models are critical in determining why moving to the cloud is important, and useful in terms of lowering the Total cost of ownership (TCO). Cloud cost calculators can be leveraged to benchmark application performance / usage / trends, which would determine in predicting overall costs of the system".

Finally, E3 explained that:

"One interesting question is whether the application demand variations are such that justify the migration. If the application usage (and resource demand) has very few peaks for example or is very stable, then it does not exploit one of the basic features of clouds with regard to resource elasticity and pay-as-you-go model".

It is quite possible that the knowledge about the legacy systems being outdated, imperfect, and undocumented. Hence, an in-depth understanding of their current state may be required for better migration outcome. The activity *Identify Legacy System* is to recapture an abstract As-Is representation of legacy systems, their functionality, dependencies to other systems, interaction points and message follows between these systems, and quality of legacies' code blocks for reuse and adaptation. Participants confirmed that the necessity of this activity by terms such as "understanding how the legacy communicates" E6, "analysing legacy dependencies" E7, and "code understanding" E3.

The activity *Analyse Migration Requirements* is to specify organisational goals and expectations that are to be met by the cloud migration. These may include computational requirements, servers, data storage and security, networking and response time, and elasticity requirements. Respondents mentioned that a requirement analysis in the context of cloud migration needs to have a focus on aspects such as "costing of elasticity support" E12 and "costs due to variable infrastructure" E8.

As name implies, the activity *Define Plan* is to specify a sequence of further activities guiding the migration process. It may include a procedure for notifying users of temporal unavailability of legacy systems, a proper roll-back plan to in-house version of legacies in the case of occurrence of any failures, and a retirement scheduling of legacies. E21 mentioned about the considering temporarily disabling access to legacy systems as a part of planning activity. He stated that

"Depends on the application, for a critical application this should happen as unnoticeable as possible, for some, this can happen over the weekend when no one is using the application".

Similarly, E15 referred to this as "downtime window for migration". Since the migration might be time-consuming process rather than instantaneous, incoming requests by other working systems to legacy systems should be passed to a mediator gateway in the course of the migration process. Such gateway by translating and redirecting calls from the legacy systems to other target systems allows other working systems to interact with legacy systems during migration.

Finally, a proper plan in order to roll back from the cloud to an in-house version of legacy system at any stages of the migration will significantly reduce the risk and exposure to the organisation. E9 used the phrase "hot swapping (roll-back planning)" to highlight this aspect of the migration planning. E8 with the experience of participating in 5 migration projects pointed out incorporating roll-back plan as a part of planning:

"In a migration methodology, there should be tasks to roll back the cloud application to a previous version if things go wrong. In a worst-case scenario, cloud application needs to pull back to be an on-premise application".

### 4.2 Design phase

The primary goal of this phase is to define an appropriate cloud architecture model for the legacy systems in line with goals and requirements defined in the *Plan Phase*. For many reasons such as data security, privacy laws, and potential performance and network latency in cloud environments, an organisation may move some components of a legacy system to cloud whilst other system components are kept in the local organisation network and cloud services are offered to them. In such situations, the activity *Design Cloud Solution* in the process model is to split a legacy system into its components and identify an optimum set of them to be migrated to the cloud regarding non-functional requirements such as data privacy, expected workload profile, and acceptable network latency. Next, an appropriate distribution and deployment of these selected components on cloud servers should be specified. Again factors such as availability zone of cloud providers, affinity of servers, acceptable network latency, and the geographical location of cloud servers should be taken into account. E10 explained how designing a cloud architecture model can become a challengeable activity:

> "As the architecture of the application becomes more complex, i.e. involving different computational nodes, storage nodes, database nodes, redundant nodes, temporal nodes to cover increases on the demands and different providers. It is more crucial to have a clear idea of where the different parts of the systems are running, or where they can run".

The activity *Choose Cloud Platform/Provider* is to identify, analyse, and prioritise cloud services that suit cloud migration objectives and legacies' requirements. Respondents mentioned several factors that should be considered when analysing cloud services available in market place. These include "data lock-in concerns" E15, "ability to offer a combination of IaaS, PaaS and SaaS to support hybrid architecture during the transition period" E4, "legacy system requirements" E13, "heterogeneity among varied cloud vendors" E11, "benchmarking cloud service offering candidates" E14, and "cloud offering selection cannot be decoupled from the design" E23.

Advancements in the cloud computing are still on an ongoing track and there is not common standard for development of cloud services [26]. As a result, integrating legacies to cloud services is a challenging process because cloud services are offered by different providers with different underlying technologies and proprietary APIs. If a legacy is to be bound to specific set of cloud services, potential incompatibilities between the legacy systems and these services should be identified and resolved accordingly through legacy modification mechanisms. This consideration is defined in the process model by the activity called *Identify Incompatibilities*. Performing this activity gives developers an estimation of required adaptation effort to enable legacies to be integrated with cloud services. E3 described the importance of this activity as follow:

> "Provider selection may produce additional requirements in terms of implementation. This refers to the fact that based on what provider (e.g. Google App engine or Azure), specific requirements may be posed on the technical implementation. For example, if you go for PaaS, then providers usually support only a subset of the available Java frameworks. If the migration is from an SQL-based database for example to a SQL based service database, then migration is easy. If migration involves the alteration of the data model (e.g. to a NoSQL solution), then significant modifications must be performed and more importantly to check the constraints of the new used paradigm. Thus you must take that under consideration, especially when you want to reutilize existing code, which is the case in application migration. Checking which frameworks are supported against which frameworks have been used for your initial application gives you an indication of how complex it would be to migrate to that specific provider".

A similar argument was reiterated by **E17** who mentioned:

> "Cloud providers will have certain limitations (e.g. which platforms are supported, which features). An example, using Azure SQL will have limitations, as not all SQL functions can be used. Therefore application might need to be adapted to support a selected cloud provider".

During the re-architecting of a legacy system to utilise cloud services, a number of design principles should be taken into account as described in the following. A notable design principle is *Handle Transient Faults* (Figure 1). Despite the high availability of cloud services, random transient faults such as temporal faults in establishing a connection to a cloud service are likely to occur. Therefore, the legacy system should be enabled by implementing mechanisms to detect and handle different transient faults in the cloud. E10 shared his opinion and explained that:

> "Providers can disappear. It is quite important to analyse the need to implement mitigation features in the application to address possible discontinuities in the related providers".

Another important design principle is *Decouple Legacy System Components*. Thanks to the proliferation of cloud servers allowing software systems leverage multiple-clouds, systems can change their deployment environment at the runtime regarding specific needs of the system, or user's preferences. In order to support dynamic deployability and independent scalability of a system, the system needs to be modified for being stateless to minimise storing contextual data during their execution. The decoupling of legacy system components, enables independent elastic scaling of the components by dynamically adding/removing more instances of the same component, transparent access to components, and coping with failures. The importance of such a design principle was stated by E24 as a way "to minimise the amount of time needed for code refactoring if changing cloud provider". Additionally, E16 stated the importance of decoupling for the fault tolerance when the system is run in the cloud. He said:

> "When you deploy on a cloud like EC2, your architecture has to take into account that parts could fail at any time: e.g., servers can go down, networks can have transient problems. In this sort of environment, you want to avoid deploying single monolithic. So what you do is break up your application into independent services, and you try to make those services stateless where you can. It makes the failure domain smaller by breaking up your application into services like this".

Coexistence of multiple instances of a system (or its components) at the runtime in the cloud requires clear session management of the system in order to track tenant's activities in the cloud, and to dynamically ensure their security when system instances are added or removed. Hence, a legacy system should have enabled implementing session management mechanisms. This is captured in the proposed model by the design principle *Make Legacy System Stateless*.

Finally, to support high business continuity and minimal downtime, the system components should be deployed over multiple cloud servers. This implies enabling the system to support synchronisation of its components (local replicas and those hosted in the cloud). To this end, the process model defines the elements *Replicate Components* and *Synchronise Components*.

### 4.3 Enable phase

This phase embraces a series of code modifications in legacy system in order to enable it to interact with and utilise cloud services in accordance with a cloud solution designed in the *Design phase*. The modifications may lead to either the implementation of new components to be integrated with the existing system components or hosted in a cloud server separate from the system.

As shown in Figure 1, the process model includes three types of modification activities for dealing with incompatibility issues between legacy systems and cloud services. These are classified under the process element *Resolve Incompatibilities*. A basic form of the modification is to *Refactor Code* which is to modify legacy system codes and interfaces to remove mismatches with cloud service. Mismatches can be in the form of interface signature, operation ordering, operation names, message format, interaction protocols, and data type. Alternatively, developing integrators (or wrappers) can serve to resolve incompatibilities. This is captured by the activity *Develop Integrator* in the process model. Integrators provide an abstraction layer, keeping legacy system codes untouched facilitating system interoperability and portability over multiple clouds. The third type of modification is captured by the element *Adapt Data* which is related to the modification of legacy data. Migrating legacy data to a cloud database solution might cause different incompatibility issues so that business logic layer cannot call data access layer functionalities as it previously did. For instance, migration to/from a relational database from/to a non-relational/non SQL database (e.g. Amazon SimpleDB) may imply applying different adaptations such as data type conversions, query transformation, database schema transformation, and implementing runtime emulators.

In cloud environments, each service consumer is called a tenant [72, 73]. Multi-tenancy is an ability to use the same instance of a cloud service at the same time by different tenants. It maximises resource utilisation and profit since only one system instance is required to deploy in the cloud. Each tenant feels that he/she is the only users of the systems and able to customise system components such as user interface appearance, business rules, sequence of workflow execution, and last but not least the system code. A major concern in the migration from single-tenant architecture to a multi-tenants one is the isolation of tenants that concurrently access the system components such as data and resources.

In the proposed process model, there are five sub-classes of the construct *Isolate Tenants* which maintain security, performance, fault, customisability, and availability of tenants using the system and cloud services (Figure 1). Each tenant should be properly authorised, load a right instance of the system, and be protected from unauthorised access by other tenants that are using the system, as E6 pointed out:

> "The big issue with multi-tenancy is the security posture of the system. Some customers don't want their data mixed with other customers' data. If you have to use a multi-tenant model then your security controls must be in place and you need to be able to explain what those controls are".

To address system security in the cloud, the superclass element *Encrypt/Decrypt Entities* (Figure 1) has three subclasses as defined in the following. The code blocks of the system which reflect an organisation's business processes might need to be secured prior to hosting in the cloud in order to protect unauthorised access by other tenants which are running on the same cloud. With respect to this, in the process model the element *Obfuscate Codes* is to assure the confidentiality of code execution through encryption mechanisms so that no other tenants will be able to access, read, or alter the code blocks within the system when running on cloud servers. In the case of deploying database in the cloud, not only the database might be threatened by external malicious tenant attacks, but also cloud providers may be granted access to the database and they may deliberately or inadvertently adventure the database [74]. Furthermore, as the cloud providers may have their subcontractors, the subcontractors may also have access to the database and affect organization's database confidentiality. With respect to this E14 said:

> "When you move your data into the cloud you lose the control who can access, manipulate, etc. it. Again for most people using Facebook, Google Services, etc. privately unfortunately it is not important and they do not care, but for enterprise applications it is highly important that you protect your data and control the access, etc. to it".

Therefore, encrypting system's database prior to hosting it in the cloud environment, as captured by the element *Encrypt Database*, is to address database confidentiality concern. Database encryption subsequently may imply extending the data access layer with the new functionalities so as to run SQL queries over encrypted database including select, join order, and aggregate operations as stated by [S43]. The third subclass of *Encrypt/Decrypt Entities* is *Message encryption/decryption* which is performed to secure messages transmission between the local legacy system components and those components which are either hosted on the cloud or distributed across multiple clouds.

Elasticity is the ability to scale up and down depending on resource demand and provide a mean for optimizing resource usage [75]. If the required resources of a legacy system significantly vary during in the course of its execution, the system should be extended to support dynamic resource acquisition and release of cloud resources. With respect to this, the element *Enable Elasticity* is defined in the process model. E22 said:

> "Considering efficient use of resources by the application needs to be taken care of in entire software development lifecycle; design, programming and coding, DB design and so because cost model is different than legacy deployments and is usually usage based in cloud-based applications and services".

Generally speaking, two common types of scalability are applied to cloud environments to provide a foundation for the elasticity [34]. There are *vertical* and *horizontal* scalability. The former is to add new computational resources to a cloud server (e.g. more CPU, RAM or hard drive space) and the latter is to add more instances of cloud servers for getting enhanced system performance. Implementation of the vertical scalability depends on the ability of the cloud service provider to support this feature. On the other hand, the horizontal scalability depends on how legacy system components are able to handle it and therefore it requires planning and ensuring whether required resources are available. For instance, re-architecting a legacy system performing transactions to support a horizontal scalability would need replicating data layers in servers and thus implementing additional concurrency management mechanisms in legacy systems.

Elasticity feature in a legacy system can be realised by implementing a new elasticity controller being embed in the system or separately hosted in the cloud server which continuously monitors the system resource usage variables and performs appropriate action to acquire or release resources on the basis of a set of scaling rules/conditions related to specific workload, threshold, events, or metrics.

Additionally, the construct *System Rebalancing* in the process model is the termination of system instances in a low-preference cloud and resuming its execution in a high-preferred cloud in order to satisfy system user preferences. If required, the legacy system should be extended to a new component which will be responsible for rebalancing the system at the runtime based on a set of rebalancing policies. **E25** shared his own experience by saying:

> "Regarding application re-balancing, typically we used various cloud providers based on their location and performance to our end users. Development was all done in the same (lowest-cost) cloud provider while production was pushed to any cloud provider that could run our applications and be close to the users".

Once modifications are applied to legacy systems through abovementioned process elements, the system will then be examined through the activity *Test System*. This element is performed to ensure that system conforms to the expectation of the cloud migration. As shown in Figure 1, the process model includes several subclasses of the test activity including *Test Network Connectivity*, *Test Scalability*, *Test Multi-Tenancy*, *Test Security*, *Test Performance*, *Test Interoperability*, *Test Scalability*, and *Test System.*

Finally, the activity *Deploy System* refers to installation of legacy components on the cloud servers. Different third-party libraries and tools (.e.g. monitoring and reporting tools) might be required to be installed both in the cloud and the local network; and this should be done prior to running the system in the cloud. Once the system migrated to the cloud, the connection between the local network and the migrated system in the cloud may still be required. In this regard, the activity *Reconfigure Network* is performed to reassign network addresses and configuring policies to assure the security of the system will be satisfied in the cloud. This will assure that tenants' reachability policies for accessing the system still are observed in the cloud.

## 5. Research Implications

Prior researches largely have been devoted to either adhocracy development of technical-oriented solutions for the integration of legacy systems to cloud services and identification of the drivers and determinants for aspects of successful cloud adoption. This article contributes to the cloud computing literature by deriving a set of important process elements involved in the moving existing legacy systems from on-premise to cloud platforms. Twenty nine elements are identified and categorized in three groups plan, design, and enable phases. The study results in a complementary model are consistent with previous research in the sense that it consists of more granular process elements discussed in ex-ante literature. They are critical to understand how to conduct legacy system reengineering to cloud platforms. Such understanding in turn is expected to lead to concrete guidelines for organizations which are either at the edge of transition to the cloud or are moving from an existing cloud platform to a new cloud platform in post-adoption phase. [5] identifies and classifies studies relevant to the cloud migration, focusing on existing techniques and methods and reflecting on areas of future research. A recent study [6] proposes an evaluation framework to highlight criteria that an ideal cloud migration process model should adhere to. The current study extends cloud computing body of knowledge by presenting a model that facilitates understanding of underlying structure of the cloud migration process by unifying and distilling cumulated knowledge in the literature evaluated by domain experts. This in turn is expected to enhance communication among various roles involved in the migration process.

This model is a useful starting point for providing contextual education to the current IS scholars and practitioners as well as newcomers to the cloud computing field to envisage an overall view of the transition to the cloud environments. From this angle, this research can be viewed as a response to the call made by Tran et al. [76] where they identified key factors affecting cloud migration costs, among them the costs involved in the learning curve of cloud migration as a cause of cost overheads. Also, due to the generic and independent nature of the presented model from any particular cloud migration scenario or underlying technology, it can also be thought as a knowledge sharing artefact for enhancing communication and knowledge transfer among IS scholars and practitioners by acting as a reference model. The need for this is corroborated by [25, 77].

Furthermore, from the perspective of information system development, the model presented in this research provides a tool for organisations to analyse and appraise their acquisition method alternatives (e.g. in-house or off-the-shelf methods) when moving from legacy to cloud platforms, as well as better understanding of the shortcomings, strengths, similarities, and differences among these alternative methods. This work further heeds Siau and Rossi recommendation for using a meta process model in IS method selection [78].

## 6. Conclusion and future work

Prior research in the cloud computing emphasises the availability of mechanisms that can ensure increased benefit and effectiveness of the cloud computing technology. To extend the trend and grounded in an extensive qualitative analysis of the literature and quantitative and qualitative evaluation by domain experts, we presented a legacy-to-cloud process model to explain how organizational legacy systems can be reengineered to utilize cloud services. Without entangling with technical implementation aspects of cloud migration, the model provides sufficiently generic insight into the key challenges in such transition. These insights include analysing organisational context, understanding cloud migration requirements, planning, understanding legacy systems, choice and impact of cloud platform for maintaining incompatibility between legacy systems and cloud services and subsequent code refactoring and data adaptation, distribution of legacy system components on cloud servers, tenant isolation, environment configuration, and testing.

One limitation of this study is that even though the model is drawn from the literature and evaluated using domain experts, we do not claim its generalisability. There are still areas to augment the model by including new elements, in particular ones related to the post migration, i.e. referring to the phase after legacy systems are successfully moved from local original environment to a target cloud platform, which have received less attention in the current literature [79]. These elements may be related to the continuous monitoring and collecting critical data about legacy system health to assure SLAs and identify any violations of the agreement from cloud provider or areas that might need special attention, billing management to track tenants that use the services and accordingly bill them for their usage of the system in terms of time and resource, and system back up, patch update, decommissioning and withdrawing cloud services, and continuous integration of the system in the cloud in a smooth and coordinated manner.

There is no universally superior or applicable method for all cloud migration scenarios and thus methods should be tailored with respect to the specific characteristics of the given scenario [80]. One option to resolve this issue is to extend the current study to provide a method engineering foundation [81, 82]. The method engineering approach is based on the idea that instead of looking for a universal information system development method, IS developers should construct methods by reusing and enhancing existing methods to meet project goals. The reuse gets supported with a model driven retrieval approach as illustrated in [83]. Reusing the actual existing process elements, which form a knowledge repository of key elements of cloud process with a possibility for extension, will enable constructing instances of migration methods that fit characteristic of a given migration scenario. By characteristics we mean factors such as code refactoring cost, the choice of a migration variant and target cloud platform, the pricing model of cloud providers, the capability of the development team, and time to market which may influence on method customisation.

## References

[1]  M. Armbrust, A. Fox, R. Griffith, A. D. Joseph, R. Katz, A. Konwinski, G. Lee, D. Patterson, A. Rabkin, and I. Stoica, "A view of cloud computing," *Communications of the ACM,* vol. 53, pp. 50-58, 2010.

[2]     R. Buyya, C. S. Yeo, and S. Venugopal, "Market-oriented cloud computing: Vision, hype, and reality for delivering it services as computing utilities," in *High Performance Computing and Communications, 2008. HPCC'08. 10th IEEE International Conference on*, 2008, pp. 5-13.

[3]     S. A. Koçak, A. Miranskyy, G. I. Alptekin, A. B. Bener, and E. Cialini, "The Impact of Improving Software Functionality on Environmental Sustainability," *on Information and Communication Technologies,* p. 95, 2013.

[4]     Gartner, *[WWW document] http://www.gartner.com/newsroom/id/3354117 (accessed 22 September 2016).* 2016.

[5]     P. Jamshidi, A. Ahmad, and C. Pahl, "Cloud Migration Research: A Systematic Review," *Cloud Computing, IEEE Transactions on,* vol. PP, pp. 1-1, 2013.

[6]     M. Fahmideh, F. Daneshgar, G. Low, and G. Beydoun, "Cloud migration process—A survey, evaluation framework, and open challenges," *Journal of Systems and Software,* vol. 120, pp. 31-69, 2016.

[7]     S. E. Chang, K.-M. Chiu, and Y.-C. Chiao, "Cloud migration: Planning guidelines and execution framework," in *Ubiquitous and Future Networks (ICUFN), 2015 Seventh International Conference on*, 2015, pp. 814-819.

[8]     Z.Mahmood, in *Cloud Computing Methods and Practical Approaches*, ed: Springer-Verlag London 2013, p. 64.

[9]     J.-F. Zhao and J.-T. Zhou, "Strategies and methods for cloud migration," *International Journal of Automation and Computing,* vol. 11, pp. 143-152, 2014.

[10]    E. A. N. da Silva and D. Lucredio, "Software Engineering for the Cloud: A Research Roadmap," in *Software Engineering (SBES), 2012 26th Brazilian Symposium on*, 2012, pp. 71-80.

[11]    Merriam-Webster, 2013.

[12]    K. Bennett, "Legacy Systems: Coping with Success," *IEEE Softw.,* vol. 12, pp. 19-23, 1995.

[13]    C. Holland and B. Light, "A critical success factors model for ERP implementation," *Software, IEEE,* vol. 16, pp. 30-36, 1999.

[14]    M. Brodie and M. Stonebraker, *Migrating legacy systems: gateways, interfaces \& the incremental approach*: Morgan Kaufmann Publishers Inc., 1995.

[15]    J. Bisbal, D. Lawless, W. Bing, and J. Grimson, "Legacy information systems: issues and directions," *Software, IEEE,* vol. 16, pp. 103-111, 1999.

[16]    R. Khadka, A. Saeidi, A. Idu, J. Hage, and S. Jansen, "Legacy to SOA Evolution: A Systematic Literature Review," in *In AD Ionita, M. Litoiu, & G. Lewis (Eds.) Migrating Legacy Applications: Challenges in Service Oriented Architecture and Cloud Computing Environments*, 2013.

[17]    H. M. Sneed, "Integrating legacy software into a service oriented architecture," in *Software Maintenance and Reengineering, 2006. CSMR 2006. Proceedings of the 10th European Conference on*, 2006, pp. 11 pp.-14.

[18]    H. M. Sneed, "Planning the reengineering of legacy systems," *Software, IEEE,* vol. 12, pp. 24-34, 1995.

[19]    L. Erlikh, "Leveraging legacy system dollars for e-business," *IT Professional,* vol. 2, pp. 17-23, 2000.

[20]    A. Bhattacherjee and S. C. Park, "Why end-users move to the cloud: a migration-theoretic analysis," *European Journal of Information Systems,* vol. 23, pp. 357-372, 2014.

[21]    S. C. Misra and A. Mondal, "Identification of a company's suitability for the adoption of cloud computing and modelling its corresponding Return on Investment," *Mathematical and Computer Modelling,* vol. 53, pp. 504-521, 2011.

[22]    A. Khajeh-Hosseini, D. Greenwood, J. W. Smith, and I. Sommerville, "The cloud adoption toolkit: supporting cloud adoption decisions in the enterprise," *Software: Practice and Experience,* vol. 42, pp. 447-465, 2012.

[23] T. Oliveira, M. Thomas, and M. Espadanal, "Assessing the determinants of cloud computing adoption: An analysis of the manufacturing and services sectors," *Information & Management,* vol. 51, pp. 497-510, 2014.

[24] H.-L. Truong and S. Dustdar, "Composable cost estimation and monitoring for computational applications in cloud computing environments," *Procedia Computer Science,* vol. 1, pp. 2175-2184, 2010.

[25] M. Hamdaqa and L. Tahvildari, "Cloud computing uncovered: a research landscape," *Advances in Computers,* vol. 86, pp. 41-85, 2012.

[26] A. N. Toosi, R. N. Calheiros, and R. Buyya, "Interconnected cloud computing environments: Challenges, taxonomy, and survey," *ACM Computing Surveys (CSUR),* vol. 47, p. 7, 2014.

[27] F. Pallas, "An agency perspective to cloud computing," in *International Conference on Grid Economics and Business Models*, 2014, pp. 36-51.

[28] O. Yigitbasioglu, "Modelling the intention to adopt cloud computing services: a transaction cost theory perspective," *Australasian Journal of Information Systems,* vol. 18, 2014.

[29] A. Menychtas, C. Santzaridou, G. Kousiouris, T. Varvarigou, L. Orue-Echevarria, J. Alonso, J. Gorronogoitia, H. Brunelière, O. Strauss, and T. Senkova, "ARTIST Methodology and Framework: A novel approach for the migration of legacy software on the Cloud," in *Symbolic and Numeric Algorithms for Scientific Computing (SYNASC), 2013 15th International Symposium on*, 2013, pp. 424-431.

[30] P. Mohagheghi, A. Berre, A. Henry, F. Barbier, and A. Sadovykh, "REMICS- REuse and Migration of Legacy Applications to Interoperable Cloud Services," in *Towards a Service-Based Internet*. vol. 6481, E. Nitto and R. Yahyapour, Eds., ed: Springer Berlin Heidelberg, 2010, pp. 195-196.

[31] M. Menzel and R. Ranjan, "CloudGenius: decision support for web server cloud migration," in *Proceedings of the 21st international conference on World Wide Web*, 2012, pp. 979-988.

[32] S. Frey and W. Hasselbring, "The cloudmig approach: Model-based migration of software systems to cloud-optimized applications," *International Journal on Advances in Software,* vol. 4, pp. 342-353, 2011.

[33] M. A. Chauhan and M. A. Babar, "Towards Process Support for Migrating Applications to Cloud Computing," in *Cloud and Service Computing (CSC), 2012 International Conference on*, 2012, pp. 80-87.

[34] V. Andrikopoulos, T. Binz, F. Leymann, and S. Strauch, "How to adapt applications for the Cloud environment," *Computing,* vol. 95, pp. 493-535, 2013/06/01 2013.

[35] C. Pahl, H. Xiong, and R. Walshe, "A Comparison of On-Premise to Cloud Migration Approaches," in *Service-Oriented and Cloud Computing*, ed: Springer, 2013, pp. 212-226.

[36] A. Fox, R. Griffith, A. Joseph, R. Katz, A. Konwinski, G. Lee, D. Patterson, A. Rabkin, and I. Stoica, "Above the clouds: A Berkeley view of cloud computing," *Dept. Electrical Eng. and Comput. Sciences, University of California, Berkeley, Rep. UCB/EECS,* vol. 28, 2009.

[37] R. Chow, P. Golle, M. Jakobsson, E. Shi, J. Staddon, R. Masuoka, and J. Molina, "Controlling data in the cloud: outsourcing computation without outsourcing control," in *Proceedings of the 2009 ACM workshop on Cloud computing security*, 2009, pp. 85-90.

[38] J. Pepitone, "Amazon EC2 outage downs Reddit, Quora," *Retrieved May,* vol. 17, p. 2011, 2011.

[39] D. Linthicum, "Why Cloud Computing Projects Fail?," *Available at: http://www.slideshare.net/Linthicum/why-cloud-computing-projects-fail, last access October 2016,* 2012.

[40] J. Tsidulko, "The 10 Biggest Cloud Outages Of 2016," *Available at http://www.crn.com/slide-shows/cloud/300081477/the-10-biggest-cloud-outages-of-2016-so-far.htm,* 2016.

[41] B. Bosker, "Users report all emails DELETED," *The Huffington PostAvailable at http://www.huffingtonpost.com.au/entry/gmail-reset-emails-deleted_n_828863, February 27 2011, Last accessed on March 4, 2017* 2011.

[42]  A. Abdollahzadegan, C. Hussin, A. Razak, M. Moshfegh Gohary, and M. Amini, "The organizational critical success factors for adopting cloud computing in SMEs," *Journal of Information Systems Research and Innovation (JISRI),* vol. 4, pp. 67-74, 2013.
[43]  S. Trigueros-Preciado, D. Pérez-González, and P. Solana-González, "Cloud computing in industrial SMEs: identification of the barriers to its adoption and effects of its application," *Electronic Markets,* vol. 23, pp. 105-114, 2013.
[44]  S. Marston, Z. Li, S. Bandyopadhyay, J. Zhang, and A. Ghalsasi, "Cloud computing—The business perspective," *Decision Support Systems,* vol. 51, pp. 176-189, 2011.
[45]  J. W. Creswell, *Research design: Qualitative, quantitative, and mixed methods approaches*: Sage publications, 2013.
[46]  X. N. Deng, T. Wang, and R. D. Galliers, "More than providing 'solutions': towards an understanding of customer-oriented citizenship behaviours of IS professionals," *Information Systems Journal,* vol. 25, pp. 489-530, 2015.
[47]  E. Tom, A. Aurum, and R. Vidgen, "An exploration of technical debt," *Journal of Systems and Software,* vol. 86, pp. 1498-1516, 2013.
[48]  G. Piccoli and B. Ives, "Trust and the unintended effects of behavior control in virtual teams," *MIS quarterly,* pp. 365-395, 2003.
[49]  S. Newell and L. F. Edelman, "Developing a dynamic project learning and cross-project learning capability: synthesizing two perspectives," *Information Systems Journal,* vol. 18, pp. 567-591, 2008.
[50]  J. Mingers, "Combining IS research methods: towards a pluralist methodology," *Information systems research,* vol. 12, pp. 240-259, 2001.
[51]  V. Venkatesh, S. A. Brown, and Y. W. Sullivan, "Guidelines for Conducting Mixed-methods Research: An Extension and Illustration," *Journal of the Association for Information Systems,* vol. 17, p. 435, 2016.
[52]  R. L. Kumar and A. C. Stylianou, "A process model for analyzing and managing flexibility in information systems," *European Journal of Information Systems,* vol. 23, pp. 151-184, 2014.
[53]  M. F. Gholami, M. Sharifi, and P. Jamshidi, "Enhancing the OPEN Process Framework with service-oriented method fragments," *Software & Systems Modeling,* pp. 1-30, 2014.
[54]  S. H. Othman and G. Beydoun, "Model-driven disaster management," *Information & Management,* vol. 50, pp. 218-228, 2013.
[55]  G. Beydoun, G. Low, B. Henderson-Sellers, H. Mouratidis, J. J. Gomez-Sanz, J. Pavon, and C. Gonzalez-Perez, "FAML: a generic metamodel for MAS development," *Software Engineering, IEEE Transactions on,* vol. 35, pp. 841-863, 2009.
[56]  G. Beydoun, C. Gonzalez-Perez, B. Henderson-Sellers, and G. Low, "Developing and evaluating a generic metamodel for MAS work products," in *International Workshop on Software Engineering for Large-Scale Multi-agent Systems*, 2005, pp. 126-142.
[57]  A. Schwarz, M. Mehta, N. Johnson, and W. W. Chin, "Understanding frameworks and reviews: a commentary to assist us in moving our field forward by analyzing our past," *ACM SIGMIS Database,* vol. 38, pp. 29-50, 2007.
[58]  S. H. Othman, G. Beydoun, and V. Sugumaran, "Development and validation of a Disaster Management Metamodel (DMM)," *Information Processing & Management,* vol. 50, pp. 235-271, 2014.
[59]  B. Kitchenham, O. Pearl Brereton, D. Budgen, M. Turner, J. Bailey, and S. Linkman, "Systematic literature reviews in software engineering – A systematic literature review," *Information and software technology,* vol. 51, pp. 7-15, 2009.
[60]  A. Khajeh-Hosseini, D. Greenwood, and I. Sommerville, "Cloud migration: A case study of migrating an enterprise it system to iaas," in *Cloud Computing (CLOUD), 2010 IEEE 3rd International Conference on*, 2010, pp. 450-457.
[61]  S. Strauch, V. Andrikopoulos, D. Karastoyanova, and K. Vukojevic, "Migrating eScience Applications to the Cloud: Methodology and Evaluation," 2014.

[62] A. Pinsonneault and K. Kraemer, "Survey research methodology in management information systems: an assessment," *Journal of management information systems,* vol. 10, pp. 75-105, 1993.
[63] J. Snow and M. Mann, "Qualtrics survey software: handbook for research professionals," ed: Obtenido de http://www.qualtrics.com, 2013.
[64] M. Q. Patton, *Qualitative evaluation and research methods*: SAGE Publications, inc, 1990.
[65] B. Kitchenham and S. L. Pfleeger, "Principles of survey research: part 5: populations and samples," *ACM SIGSOFT Software Engineering Notes,* vol. 27, pp. 17-20, 2002.
[66] G. A. Morgan, N. L. Leech, G. W. Gloeckner, and K. C. Barrett, *SPSS for introductory statistics: Use and interpretation*: Psychology Press, 2004.
[67] H. W. Lilliefors, "On the Kolmogorov-Smirnov test for normality with mean and variance unknown," *Journal of the American Statistical Association,* vol. 62, pp. 399-402, 1967.
[68] M. Rosenblatt, "A central limit theorem and a strong mixing condition," *Proceedings of the National Academy of Sciences of the United States of America,* vol. 42, p. 43, 1956.
[69] G. V. Glass, P. D. Peckham, and J. R. Sanders, "Consequences of failure to meet assumptions underlying the fixed effects analyses of variance and covariance," *Review of educational research,* pp. 237-288, 1972.
[70] L. M. Lix, J. C. Keselman, and H. Keselman, "Consequences of assumption violations revisited: A quantitative review of alternatives to the one-way analysis of variance F test," *Review of educational research,* vol. 66, pp. 579-619, 1996.
[71] A. Field, *Discovering statistics using IBM SPSS statistics*: Sage, 2013.
[72] B. P. Rimal, C. Eunmi, and I. Lumb, "A Taxonomy and Survey of Cloud Computing Systems," in *INC, IMS and IDC, 2009. NCM '09. Fifth International Joint Conference on*, 2009, pp. 44-51.
[73] C. J. Guo, W. Sun, Y. Huang, Z. H. Wang, and B. Gao, "A framework for native multi-tenancy application development and management," in *E-Commerce Technology and the 4th IEEE International Conference on Enterprise Computing, E-Commerce, and E-Services, 2007. CEC/EEE 2007. The 9th IEEE International Conference on*, 2007, pp. 551-558.
[74] M. D. Ryan, "Cloud computing security: The scientific challenge, and a survey of solutions," *Journal of Systems and Software,* 2013.
[75] L. Badger, T. Grance, R. Patt-Corner, and J. Voas, "Cloud computing synopsis and recommendations," *NIST special publication,* vol. 800, p. 146, 2012.
[76] V. Tran, J. Keung, A. Liu, and A. Fekete, "Application migration to cloud: a taxonomy of critical factors," in *Proceedings of the 2nd International Workshop on Software Engineering for Cloud Computing*, 2011, pp. 22-28.
[77] O. Zimmermann, C. Miksovic, and J. M. Küster, "Reference architecture, metamodel, and modeling principles for architectural knowledge management in information technology services," *Journal of Systems and Software,* vol. 85, pp. 2014-2033, 2012.
[78] K. Siau and M. Rossi, "Evaluation of information modeling methods-a review," in *System Sciences, 1998., Proceedings of the Thirty-First Hawaii International Conference on*, 1998, pp. 314-322.
[79] D. Schlagwein and A. Thorogood, "MARRIED FOR LIFE? A CLOUD COMPUTING CLIENT-PROVIDER RELATIONSHIP CONTINUANCE MODEL," 2014.
[80] M. Fahmideh, F. Daneshgar, and F. Rabhi, "Cloud Computing Adoption: An Effective Tailoring Approach," 2016.
[81] A. F. Harmsen, J. Brinkkemper, and J. H. Oei, *Situational method engineering for information system project approaches*: University of Twente, Department of Computer Science, 1994.
[82] B. Henderson-Sellers, "Method Engineering: Theory and Practice," in *ISTA*, 2006, pp. 13-23.
[83] G. Beydoun, G. Low, F. García-Sánchez, R. Valencia-García, and R. Martínez-Béjar, "Identification of ontologies to support information systems development," *Information Systems,* vol. 46, pp. 45-60, 2014.

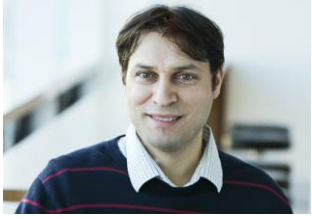
Mahdi Fahmideh Gholami completed a PhD degree in Information Systems from UNSW Business School, University of New South Wales, Sydney, Australia. His general research interests are design science research, conceptual modeling, method engineering, cloud computing, and information system development methods. He has served as a system analyst and programmer in national government IT projects for several years.

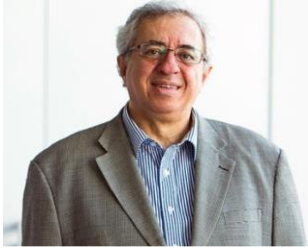
Farhad Daneshgar received his PhD in Information Systems from the University of Technology, Sydney Australia. He is a Senior Lecturer at the UNSW Business School, University of New South Wales, Sydney, Australia, and is an adjunct Professor at Bangkok University. Farhad is the creator of the Awareness Modeling Language, and has published extensively in the areas of Knowledge Management and Enterprise Systems, and was awarded twice for his Outstanding Research Article at the University of New South Wales. Farhad is a member of editorial board in five academic journals.

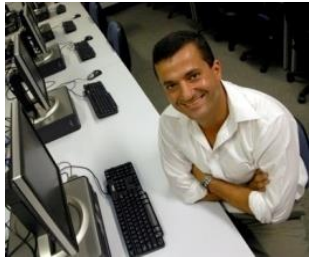
Professor Ghassan Beydoun received a degree in computer science and a PhD degree in knowledge systems from the University of New South Wales. He is currently a Professor of Information Systems at the University of Technology Sydney. He has authored more than 100 papers international journals and conferences. He is currently working on the metamodels for on project sponsored by Australian Research Council and Australian companies to investigate the endowing methodologies for distributed intelligent systems and supply chains with intelligence. His other research interests include multi agent systems applications, ontologies and their applications, and knowledge acquisition.

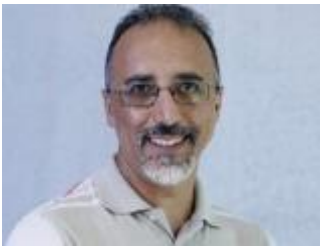
Fethi Rabhi is a Professor in the School of Computer Science and Engineering at the University of New South Wales (UNSW) in Australia. His main research areas are in service-oriented software engineering with a strong focus on business and financial applications. He completed a PhD in Computer Science at the University of Sheffield in 1990 and held several academic appointments in the USA and the UK before joining UNSW in 2000. He is currently actively involved in several research projects in the area of large-scale news and financial market data analysis.